\documentclass[preprintnumbers,10pt,nofootinbib]{revtex4}

\usepackage{amsmath,latexsym,amssymb,amsfonts}
\usepackage{graphicx}
\usepackage{bm}
\usepackage{epstopdf}
\usepackage{epsfig}
\usepackage{color}

\begin{document}

\title{\textbf{Curved accretion disks around rotating black holes without reflection symmetry}}

\author{
{\textsc{Che-Yu Chen$^{a}$}\footnote{{\tt b97202056{}@{}gmail.com}}} and
{\textsc{Hsiang-Yi Karen Yang$^{bcd}$}\footnote{{\tt hyang{}@{}phys.nthu.edu.tw}}}
}

\affiliation{
$^{a}$\small{Institute of Physics, Academia Sinica, Taipei 11529, Taiwan}\\
$^{b}$\small{Institute of Astronomy and Department of Physics, National Tsing Hua University, Hsinchu 30013, Taiwan}\\
$^{c}$\small{Center for Informatics and Computation in Astronomy, National Tsing Hua University, Hsinchu 30013, Taiwan}\\
$^{d}$\small{Physics Division, National Center for Theoretical Sciences, Taipei 10617, Taiwan}
}

\begin{abstract}
Rotating black holes without equatorial reflection symmetry can naturally arise in effective low-energy theories of fundamental quantum gravity, in particular, when parity-violating interactions are introduced. Adopting a theory-agnostic approach and considering a recently proposed Kerr-like black hole model, we investigate the structure and properties of accretion disk around a rotating black hole without reflection symmetry. In the absence of reflection symmetry, the accretion disk is in general a curved surface in shape, rather than a flat disk lying on the equatorial plane. Furthermore, the parameter $\epsilon$ that controls the reflection asymmetry would shrink the size of the prograde innermost stable circular orbits, and enhance the efficiency of the black hole in converting rest-mass energy to radiation during accretion. The retrograde innermost stable circular orbits are stretched but the effects are substantially suppressed. In addition, we find that spin measurements based on the gravitational redshift observations of the disk, assuming a Kerr geometry, may overestimate the true spin values if the central object is actually a Kerr-like black hole with conspicuous equatorial reflection asymmetry. The qualitative results that the accretion disk becomes curved and the prograde innermost stable circular orbits are shrunk turn out to be generic in our model when the reflection asymmetry is small.
\end{abstract}

\maketitle

\newpage

\tableofcontents

\section{Introduction}

General Relativity (GR) predicts the existence of black holes, which are defined within a portion of space area from which nothing can escape, including light. In spite of its extreme environment, black hole spacetime in GR is very simple in the sense that it respects the no-hair theorem. For an isolated astrophysical black hole in GR, the spacetime can be completely described by the Kerr metric and it only contains two physical observables: the mass and the spin. With the rapid developments of observational technology based on electromagnetic and gravitational waves, it is believed that in the near future, we can gradually uncover these extremely mysterious objects in the universe. In particular, by testing the no-hair theorem or the Kerr hypothesis, eventually, we can even verify whether our current understanding about black holes, which is based on GR, is still correct or not down to the horizon scale \cite{Johannsen:2010xs,Johannsen:2010ru,Johannsen:2010bi,Bambi:2011mj,Johannsen:2012ng,Yagi:2016jml,Johannsen:2016uoh,Barack:2018yly,Bambi:2015kza,Isi:2019aib}.

Completely isolated black holes can never be detected via any observations. Black holes can be detected only when they are interacting with their surroundings, such as with their companions or with the materials around the black holes. In astrophysical environments, infalling materials typically possess non-zero angular momentum and form accretion disks around the black holes. In the accretion disks, materials lose their angular momentum via viscous dissipation and spiral inward, while efficiently converting their rest-mass energy into radiation. The bright emission from the accretion disks has paramount astrophysical importance, as it has led to the discovery of X-ray binaries and active galactic nuclei, and allowed us to explore phenomena in the vicinity of the black holes that are not yet resolvable given current technology. For instance, the two commonly used methods for measuring black hole spins\footnote{Black hole spins could also be constrained by the waveforms of gravitational waves, though this method could only probe the population of black hole binaries and is more sensitive to combinations of the spin parameters of the two merging black holes.}, namely, the continuum fitting method \cite{Shafee:2006,McClintock:2006} and X-ray reflection method \cite{Fabian:1989,Brenneman:2006,Risaliti:2013}, both rely on the knowledge about accretion disk properties and spacetime geometry near the event horizon. By assuming that the accretion disk is well approximated by the standard thin-disk model \cite{Shakura:1973,Novikov:1973}, and that the inner edge of the accretion disk is defined by the innermost stable circular orbit (ISCO), one could fit the observed disk spectrum (hence called ``continuum fitting") and infer the radius of the ISCO and thus the black hole spin. Another way to measure the black hole spin is by observing the X-ray Fe $K_\alpha$ line at 6.4 keV coming from the reprocessed, or ``reflected", light off the accretion disk. The line shape is sensitive to the gravitational redshift effect due to spacetime distortions near the black hole, thereby allowing constraints on the black hole spin. The spin measurements made so far have relied on the implicit assumption of Kerr spacetime. However, if there are modifications of GR, the different metric could have non-negligible influence on the spacetime and accretion disk properties. For example, the accretion disks around black holes beyond GR have been extensively studied in the literature, such as the models in modified theories of gravity \cite{Pun:2008ae,Harko:2009rp,Harko:2009kj,Harko:2010ua,Perez:2012bx,Soroushfar:2020kgb,Liu:2020vkh,Heydari-Fard:2021ljh,Liu:2021yev}, hairy black holes \cite{Heydari-Fard:2020ugv,Collodel:2021gxu}, theories with higher dimensions \cite{Pun:2008ua,Chen:2011wb}, solutions with naked singularities \cite{Kovacs:2010xm,Shahidi:2020bla}, and in parameterized non-Kerr models \cite{Bambi:2011jq,Chen:2011rx,Bambi:2012tg}. All these considerations could potentially affect interpretations from astrophysical observations.

There are several assumptions that are commonly made to model astrophysical black holes. Some of them are stationarity, axisymmetry, asymptotic flatness, and equatorial reflection symmetry (or $\mathbb{Z}_2$ symmetry). Indeed, these are reasonable assumptions when the black hole under consideration is sufficiently isolated in the sense that the back reaction of the surrounding materials on the black hole spacetime can be fairly neglected. Technically, these assumptions also reduce the complexity in the investigation of the black hole under consideration. Even if one goes beyond GR and wants to construct a novel black hole model to test the Kerr hypothesis, the aforementioned assumptions still seem reasonable \cite{Delgado:2021jxd}. Following this line, several non-Kerr spacetimes have been proposed using different parameterization approaches \cite{Glampedakis:2005cf,Johannsen:2011dh,Johannsen:2013szh,Cardoso:2014rha,Konoplya:2016jvv,Ghasemi-Nodehi:2016wao,Konoplya:2018arm,Carson:2020dez,Chen:2019jbs}. However, interesting new physics may be hidden behind these assumptions and can never be noticed unless some of them are relaxed. One particular example, which is also the main theme in this paper, is the possibility of relaxing the equatorial reflection symmetry. The question is: What would happen if the black hole does not look symmetric with respect to the equatorial plane?

In fact, black hole solutions without $\mathbb{Z}_2$ symmetry could naturally appear in some effective theories of quantum gravity. In the effective field theory approach, one introduces an energy cutoff scale, which can be below the Planck scale, then expands the theory with respect to this cutoff scale. After truncating the expansion at some order, the gravitational Lagrangian of the effective field theory consists of the usual Einstein-Hilbert term and a series of higher-order curvature invariants. At the level of field equations, these higher-order curvature terms appear in the form of higher-derivative interactions. In particular, if the theory contains parity-violating interactions, which are constructed by the dual Riemann tensor:
\begin{equation}
\tilde{R}_{\mu\nu\alpha\beta}\equiv\frac{1}{2}\epsilon_{\mu\nu\rho\sigma}{R^{\rho\sigma}}_{\alpha\beta}\,,
\end{equation}     
the rotating black hole solutions in the theory are likely to be $\mathbb{Z}_2$ asymmetric. More explicitly, the parity-violating terms would be coupled with the black hole spin, and therefore, the $\mathbb{Z}_2$ asymmetry would be induced by the spin of the black hole. In Ref.~\cite{Cano:2019ore}, rotating black hole solutions without reflection symmetry have been obtained perturbatively in an effective field theory in which the Chern-Simons term and the Gauss-Bonnet term are coupled together through a dynamical scalar field. In addition, the violation of $\mathbb{Z}_2$ symmetry has been found in the theory containing higher-order curvature invariants constructed by $\tilde{R}_{\mu\nu\alpha\beta}$ \cite{Cardoso:2018ptl}. Due to the complexity of the theories, the rotating black hole solutions do not have analytic expressions and they can only be studied using numerical or perturbative approaches \cite{Cardoso:2019mqo,Reall:2019sah,McManus:2019ulj,Sennett:2019bpc,Cano:2020cao,Hatsuda:2020egs}. See Refs.~\cite{Cunha:2018uzc,Bah:2021jno} for other possible constructions of $\mathbb{Z}_2$ asymmetric black holes.

In this paper, we adopt a theory-agnostic approach and consider a Kerr-like black hole model to investigate the astrophysical implications of $\mathbb{Z}_2$ asymmetry. The Kerr-like black hole model was recently proposed in Ref.~\cite{Chen:2020aix} and it is characterized by the equatorial reflection asymmetry. The spacetime is asymptotically flat and reduces to Kerr spacetime at spatial infinity. In addition, the metric has a very simple analytic expression and the geodesic equations are completely separable. The metric contains a deviation function that controls the $\mathbb{Z}_2$ asymmetry. If the deviation function is turned off, the Kerr metric is recovered. In this work, we will focus on the property and structure of the accretion disk around this Kerr-like black hole. We will show explicitly that, in the absence of reflection symmetry, the circular orbits of massive particles are not confined on the standard equatorial plane, instead, each of them is lying on a plane parallel to, but different from, the equatorial plane{\footnote{Similar results have also been found in black hole solutions with NUT charge \cite{Jefremov:2016dpi,Mukherjee:2018dmm,Chakraborty:2019rna}. The NUT charge breaks the $\mathbb{Z}_2$ symmetry and the spacetime is in general asymptotically non-flat.}}. As a result, the accretion disk becomes a curved surface, instead of a flat disk confined on the equatorial plane. We will then investigate the effects of reflection asymmetry on some important quantities and observables, such as the locations of the ISCO, the radiative efficiency of the black hole accretion disk in converting rest-mass to radiation energy, and the gravitational redshifts of photons emitted by the accretion disk. The astrophysical implications of these results will be discussed.

The paper is organized as follows: In Sec.~\ref{sec:th}, we briefly review the setup of the Kerr-like metric that we are going to consider in this work, and then we discuss its spacetime structures. In Sec.~\ref{sec.geocir}, we investigate the geodesic equations of massive particles moving around the Kerr-like black hole, and then focus on the circular orbits, the ISCO, and the accretion disk around the black hole. We discuss the astrophysical implications of this spacetime in Sec.~\ref{sec.astr}, such as the radiative efficiency of the accretion disk and the gravitational redshift effect. We finally conclude in Sec.~\ref{sec.conclusion}.

\section{\label{sec:th}The Kerr-like metric without reflection symmetry}
In this section, we will briefly review the rotating black hole spacetime that we are going to consider throughout this paper. In Ref.~\cite{Chen:2020aix}, a class of Kerr-like black hole was proposed as a phenomenological model to describe a particular type of deviations from the Kerr metric. The metric of this Kerr-like spacetime contains some deviation functions (or parameters) that quantify the deviations from the Kerr metric. In particular, in the presence of the deviation functions, the spacetime is not symmetric with respect to the ``standard" equatorial plane and the $\mathbb{Z}_2$ symmetry of the spacetime is generically broken. 

In fact, there are several parameterized non-Kerr metrics proposed in the literature \cite{Glampedakis:2005cf,Johannsen:2011dh,Johannsen:2013szh,Cardoso:2014rha,Konoplya:2016jvv,Ghasemi-Nodehi:2016wao,Konoplya:2018arm,Carson:2020dez,Chen:2019jbs}, on top of which one may include additional terms that break the $\mathbb{Z}_2$ symmetry phenomenologically. One can further classify these parameterized models according to the existence of the Carter-like constant, or, in other words, the existence of hidden symmetry that is associated with the Killing tensors. This hidden symmetry can be broken in general \cite{Ghasemi-Nodehi:2016wao,Konoplya:2016jvv,Cardoso:2014rha} and could result in interesting phenomenologies \cite{Destounis:2020kss,Destounis:2021mqv}. In our setup, we will, however, assume that such hidden symmetry exists, and there exists a Carter-like constant, giving rise to the separability of geodesic equations. This assumption allows us to purely focus on the phenomenological effects coming from the $\mathbb{Z}_2$ asymmetry of spacetime. To be more explicit, whenever the model predicts any differences from Kerr results, we wish to directly conclude that these differences are coming from the violation of $\mathbb{Z}_2$ symmetry, not possibly from the violation of the Carter symmetry or something else. To this end, we shall consider the most general stationary and axisymmetric spacetime that preserves this hidden symmetry.

The construction of the Kerr-like metric \cite{Chen:2020aix} starts with the assumption that a Carter-like constant in the spacetime is preserved, which implies that the geodesic equations are completely separable. A general expression of a stationary and axisymmetric spacetime preserving this property in four dimensions was proposed by Papadopoulos and Kokkotas \cite{Papadopoulos:2018nvd} (the PK metric). In the Boyer-Lindquist coordinate system ($t,r,y,\varphi$) where $y\equiv\cos\theta$, the contravariant form of the PK metric reads \cite{Papadopoulos:2018nvd}
\begin{align}
g^{tt}&=\frac{\mathcal{A}_5(r)+\mathcal{B}_5(y)}{\mathcal{A}_1(r)+\mathcal{B}_1(y)}\,,\qquad g^{t\varphi}=\frac{\mathcal{A}_4(r)+\mathcal{B}_4(y)}{\mathcal{A}_1(r)+\mathcal{B}_1(y)}\,,\nonumber\\
g^{\varphi\varphi}&=\frac{\mathcal{A}_3(r)+\mathcal{B}_3(y)}{\mathcal{A}_1(r)+\mathcal{B}_1(y)}\,,\qquad g^{yy}=\frac{\mathcal{B}_2(y)}{\mathcal{A}_1(r)+\mathcal{B}_1(y)}\,,\nonumber\\
g^{rr}&=\frac{\mathcal{A}_2(r)}{\mathcal{A}_1(r)+\mathcal{B}_1(y)}\,.\label{PKmetric}
\end{align}
As one can see, the PK metric contains five radial functions $\mathcal{A}_i(r)$ and five polar functions $\mathcal{B}_i(y)$. By assuming different expressions for these functions, one can construct rotating black hole metrics with interesting features \cite{Destounis:2020kss,Destounis:2021mqv}. In Ref.~\cite{Chen:2020aix}, the Kerr-like metric was constructed from the PK metric by assuming the radial functions to be given by their Kerr expressions, and the polar functions are given by Kerrian forms added by some deviation functions $\tilde\epsilon_i(y)$:
\begin{equation}
\mathcal{A}_i(r)=\mathcal{A}_{i,\textrm{Kerr}}(r)\,,\qquad \mathcal{B}_i(y)=\mathcal{B}_{i,\textrm{Kerr}}(y)+\tilde\epsilon_i(y)\,.
\end{equation}
More explicitly, the radial functions are expressed as
\begin{align}
\mathcal{A}_1&=r^2\,,\quad \mathcal{A}_2=\Delta\,,\quad \mathcal{A}_3=-\frac{a^2}{\Delta}\,,\nonumber\\
\mathcal{A}_4&=-\frac{a\left(r^2+a^2\right)}{\Delta}\,,\qquad\mathcal{A}_5=-\frac{\left(r^2+a^2\right)^2}{\Delta}\,,\label{Kerrianmetric}
\end{align} 
where $\Delta\equiv r^2-2Mr+a^2$. Here, $M$ and $a$ represent the mass and the spin of the black hole, respectively. 

On the other hand, the polar metric functions $\mathcal{B}_i(y)$ have to be chosen properly such that the spacetime still recovers the Kerr metric at a very far distance away from the black hole. The most general choice for the polar metric functions that fulfills this requirement is
\cite{Chen:2020aix}:
\begin{align}
&\mathcal{B}_1=a^2y^2+\tilde\epsilon_1(y)\,,\qquad \mathcal{B}_2=1-y^2\,,\qquad\mathcal{B}_3=\frac{1}{1-y^2}\,,\nonumber\\
&\mathcal{B}_4=a\,,\qquad \mathcal{B}_5=a^2(1-y^2)+\tilde\epsilon_5(y)\,.
\end{align}
Note that only $\tilde\epsilon_1(y)$ and $\tilde\epsilon_5(y)$ are left after imposing the condition that the metric should reduce to the Kerr metric at the asymptotic region. These deviation functions can generically break the $\mathbb{Z}_2$ symmetry of the spacetime, as long as at least one of the two functions $\tilde\epsilon_1(y)$ and $\tilde\epsilon_5(y)$ is not even under $y\leftrightarrow -y$ exchange.

Furthermore, the Solar System test puts additional constraints on the model. The observational requirements that the post-Newtonian parameters $\beta$ and $\gamma$ should be very close to one imply the following constraint
\begin{equation}
\tilde\epsilon_1(y)+\tilde\epsilon_5(y)\ll M^2\,\qquad\textrm{for $|y|\le1$}\,.
\end{equation} 
In this work, we shall assume that $\tilde\epsilon_1(y)=-\tilde\epsilon_5(y)=\tilde\epsilon(y)$, such that the Solar System constraint is satisfied. Finally, given the fact that the $\mathbb{Z}_2$ asymmetry is usually induced by the spin of the black hole in effective theory frameworks, we shall consider the following parameterization for the deviation function:
\begin{equation}
\tilde\epsilon(y)=\epsilon M^2\left(\frac{a}{M}\right)^ny\,,\label{deviationepsilon}
\end{equation}
where $\epsilon$ and $n$ are two dimensionless parameters and they are supposed to be constrained by observations. Essentially, the deviation parameter $\epsilon$ quantifies the amount of the $\mathbb{Z}_2$ asymmetry in the spacetime. On the other hand, the index $n$ quantifies the sensitivity of the $\mathbb{Z}_2$ asymmetry with respect to the change of the black hole spin. The larger the index $n$ is, the more sensitive the asymmetry is with respect to the spin. In order for the $\mathbb{Z}_2$ asymmetry to be induced by the spin, one shall choose a positive index $n>0$. In the original model proposed in Ref.~\cite{Chen:2020aix}, only the case with $n=0$ was considered. Here we introduce the index $n$ to quantify the theoretical requirement that the $\mathbb{Z}_2$ asymmetry is in general induced by the spin of the black hole.{\footnote{Note that our phenomenological model of the Kerr-like metric with $n>0$ has stronger $\mathbb{Z}_2$ asymmetry for larger spins. The model is not able to capture the features of some modified gravitational theories in which the $\mathbb{Z}_2$ symmetry of black holes can be partially restored in the near extremal spin limit \cite{Doneva:2021dcc}.}

After taking account of the above theoretical and observational requirements, the Kerr-like metric in its covariant form can be written as
\begin{align}
g_{tt}&=-1+\frac{2Mr\left(r^2+a^2y^2\right)}{\left(r^2+a^2y^2\right)^2+\left(r^2-2Mr+a^2y^2\right)\tilde\epsilon(y)}\,,\label{gtt}\\
g_{\varphi\varphi}&=\frac{\left(1-y^2\right)\left(r^2+a^2y^2+\tilde\epsilon(y)\right)\left[r^4+a^4y^2+r^2\left(a^2+a^2y^2+\tilde\epsilon(y)\right)+a^2\tilde\epsilon(y)+2Mr\left(a^2-a^2y^2-\tilde\epsilon(y)\right)\right]}{\left(r^2+a^2y^2\right)^2+\left(r^2-2Mr+a^2y^2\right)\tilde\epsilon(y)}\,,\label{gphiphi}\\
g_{t\varphi}&=-\frac{2Mra\left(1-y^2\right)\left(r^2+a^2y^2+\tilde\epsilon(y)\right)}{\left(r^2+a^2y^2\right)^2+\left(r^2-2Mr+a^2y^2\right)\tilde\epsilon(y)}\,,\label{gtphi}\\
g_{rr}&=\frac{r^2+a^2y^2+\tilde\epsilon(y)}{r^2-2Mr+a^2}\,,\qquad g_{yy}=\frac{r^2+a^2y^2+\tilde\epsilon(y)}{1-y^2}\,,\label{grryy}
\end{align}
where the deviation function $\tilde\epsilon(y)$ is given by Eq.~\eqref{deviationepsilon}. It should be emphasized that our work is based on a theory-agnostic approach. The rotating spacetime that we are considering here is not an exact solution to any known modified theories of gravity in the literature. In fact, it is possible to obtain $\mathbb{Z}_2$ asymmetric black hole solutions in theories beyond GR \cite{Cano:2019ore,Cardoso:2018ptl,Cardoso:2019mqo,Reall:2019sah,McManus:2019ulj,Sennett:2019bpc,Cano:2020cao,Hatsuda:2020egs,Cunha:2018uzc,Bah:2021jno}, as mentioned in the introduction. Some of them do not necessarily have Carter symmetry \cite{Cano:2019ore,Cardoso:2018ptl,Bah:2021jno}, hence our model is not able to capture possible features arising from the loss of that symmetry. However, for those models that possess Carter symmetry and separable geodesic equations \cite{Cunha:2018uzc}, our metric is expected to be a good approximant of them. In addition, the choice of the deviation function \eqref{deviationepsilon} has the features that larger reflection asymmetry is associated with a larger deviation parameter $\epsilon$, and a more rapid black hole spin (if $n>0$), which are shared by many $\mathbb{Z}_2$ asymmetric solutions in modified theories of gravity \cite{Cano:2019ore,Cardoso:2018ptl,Bah:2021jno} (see Ref.~\cite{Doneva:2021dcc} for a counterexample). Although our model has its own limitation as mimicking real solutions in modified theories of gravity, however, due to its simplicity and capability to capture the feature of reflection asymmetry, this model perfectly serves our phenomenological purpose in this work.

There are some important hypersurfaces in this spacetime that deserve individual discussions. They are the event horizon, ergosurface, and the curvature singularity.
\begin{enumerate}
\item The event horizon: \\
In Ref.~\cite{Chen:2020aix}, it has been shown explicitly that the event horizon of the Kerr-like spacetime is determined by the roots of the equation $\Delta=0$. For astrophysical purposes, we will only focus on the outermost root. One can indeed prove that the hypersurface given by the root is a Killing horizon and a null surface \cite{Chen:2020aix}. Because the function $\Delta$ is exactly the same as its Kerr counterpart, the expression of the event horizon in the Boyer-Lindquist coordinate system, i.e., $r=r_h$, is the same as the Kerr result:
\begin{equation}
r_h=M+\sqrt{M^2-a^2}\,,\label{rh}
\end{equation}
and is independent of the deviation function $\tilde\epsilon(y)$. One should notice that the expression of the event horizon in the Boyer-Lindquist coordinate system being identical to its Kerr result does not mean that the spacetime structure of the event horizon is blind to the deviation function. In fact, to visualize the event horizon structure, one can use the isometric embedding to map the horizon geometry to a 3-dimensional Euclidean space. This has been done in Ref.~\cite{Chen:2020aix} and the $\mathbb{Z}_2$ asymmetry of the horizon shape can be explicitly shown (see Figure 1 in Ref.~\cite{Chen:2020aix}). Finally, according to the expression \eqref{rh}, we will only focus on the following range of the black hole spin: $ |a|/M\le1$, in the rest of the paper.
\item The ergosurface:\\
Similar to the Kerr spacetime, the rotation of the Kerr-like metric considered here also induces frame-dragging effects around the black hole. When an observer is moving too close to the black hole, it may enter the ergosphere in which the observer starts to corotate with the black hole and is unable to appear stationary with respect to spatial infinity. The boundary of this ergosphere is defined as the ergosurface, which, in the Boyer-Lindquist coordinate system, is determined by $g_{tt}=0$. According to Eq.~\eqref{gtt}, there are two possibilities: 
\begin{equation}
r^2-2Mr+a^2y^2=0\qquad\textrm{or}\qquad r^2+a^2y^2+\tilde\epsilon(y)=0\,.\label{gtteq0}
\end{equation}
We will show later that the surface given by the second equation in Eq.~\eqref{gtteq0} is actually a curvature singularity. In most cases, the curvature singularity is inside the event horizon and the ergosurface is given by the first equation of Eq.~\eqref{gtteq0}. Again, the structure of the ergosurface can be visualized using the isometric embedding method. This has been done in Ref.~\cite{Chen:2020aix} and the $\mathbb{Z}_2$ asymmetry can be clearly identified (see Figure 2 of that paper). As we will illustrate later, in some regions of the parameter space, the curvature singularity may appear outside the event horizon, or even outside the ergosurface. In these cases, the spacetime contains a naked singularity. We shall come back to these interesting cases shortly later. 
\item Singularity:\\
In the Kerr-like spacetime, the surface given by $r^2+a^2y^2+\tilde\epsilon(y)=0$ is a curvature singularity. In fact, on this surface, all the metric components \eqref{gtt}, \eqref{gphiphi}, \eqref{gtphi}, and \eqref{grryy} vanish. We have shown that near this surface, the Ricci scalar $R$ diverges as
\begin{equation}
R\approx\left[r^2+a^2y^2+\tilde\epsilon(y)\right]^{-3}\,,
\end{equation}
and the Kretschmann scalar $K$ also diverges. It should be mentioned that when the deviation function is zero, the Kerr metric is recovered, whose Ricci scalar is identically zero, while the Kretschmann scalar diverges inside the horizon. Indeed, the Kerr metric has a ring singularity, which is determined by $r^2+a^2y^2=0$.

In fact, depending on the values of $a$ and the parameters in the deviation function $\tilde\epsilon(y)$, the curvature singularity may appear outside the event horizon. The intersection of the surface corresponding to the event horizon and that corresponding to the naked singularity is determined by the following coupled equations:
\begin{equation}
r^2-2Mr+a^2=0\qquad\textrm{and}\qquad r^2+a^2y^2+\tilde\epsilon(y)=0\,.
\end{equation}
To be more explicit, we consider the parameterization \eqref{deviationepsilon} and assume $n=1$, that is, $\tilde\epsilon(y)=\epsilon May$ as an example. We fix the value of spin $a$ and increase the value of $|\epsilon|$ from zero. Then, we find that when $|\epsilon|$ reaches a critical value $\epsilon_{\textrm{sin}}$, the naked singularity starts to appear at one of the poles. This critical value $\epsilon_{\textrm{sin}}$ depends on the spin and can be written explicitly as
\begin{equation}
\epsilon_{\textrm{sin}}\equiv\frac{2+2\sqrt{1-a^2/M^2}}{|a|/M}\,.\label{epsiloncritical}
\end{equation}
In the extremal case, i.e., $|a|=M$, we have $\epsilon_{\textrm{sin}}=2$. Therefore, the Kerr-like spacetime with $|\epsilon|<2$ is everywhere regular outside the event horizon for all spins $|a|\le M$. On the other hand, in the non-rotating case, the spacetime reduces to the Schwarzschild metric and the naked singularity never appears. Furthermore, when one keeps increasing $|\epsilon|$ in the region $|\epsilon|>\epsilon_{\textrm{sin}}$, the region of the naked singularity increases and it may cover partially the ergosurface or even some orbits of the moving particles around the black hole. We will discuss about this in more details in the next section.
\end{enumerate}

Before closing this section, we use FIG.~\ref{fig.surface} to summarize the relative locations of different hypersurfaces mentioned above. In this figure, we consider the deviation function $\tilde\epsilon(y)=\epsilon May$ and assume $a/M=0.9$. On the $(r,y)$ plane, the outer event horizon and the ergosurface are shown by the vertical black line and the blue curve, respectively. The dashed red curves, from left to right, correspond to $\epsilon=-1$, $\epsilon=-\epsilon_\textrm{sin}\approx-3.2$, and $\epsilon=-5$, respectively. The intersection between the dashed red curve and the black line indicates the appearance of naked singularities. Notice that when $|\epsilon|=\epsilon_\textrm{sin}$, the naked singularity starts to appear at one of the poles.

\begin{figure}[t!]
\centering
\includegraphics[angle =0,scale=0.55]{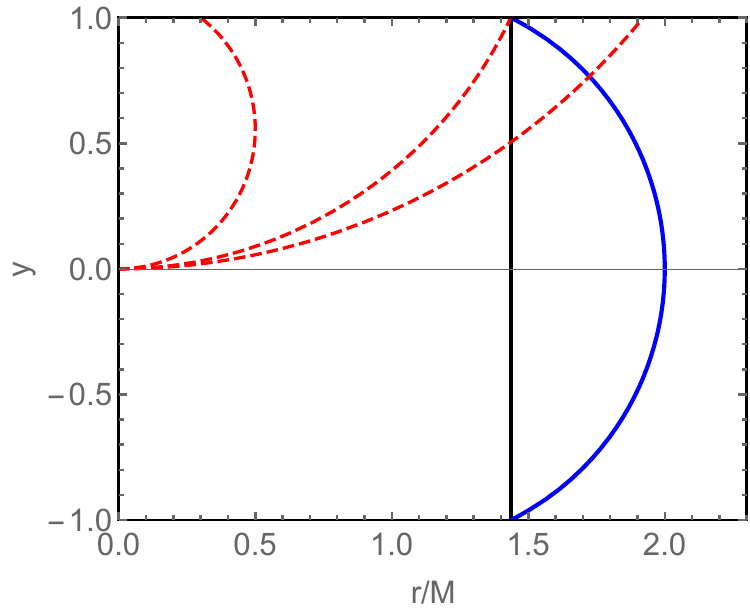}
\caption{Different hypersurfaces in the Kerr-like spacetime are depicted on the $(r,y)$ plane. We assume $a/M=0.9$. The outer event horizon and the ergosurface are uniquely determined by $a/M$ and they are shown by the vertical black line and the blue curve, respectively. The dashed red curves represent the curvature singularity. We consider the deviation function $\tilde\epsilon(y)=\epsilon May$. The red curves from left to right correspond to $\epsilon=-1$, $\epsilon=-\epsilon_\textrm{sin}\approx-3.2$, and $\epsilon=-5$, respectively. The intersection between the dashed red curve and the black line indicates the appearance of naked singularties.}
\label{fig.surface}
\end{figure}

\section{Geodesic equations, circular motions, and ISCO}\label{sec.geocir}
In order to construct the model describing the accretion disk, we will start with the geodesic equations and the circular motions of materials around the Kerr-like black hole. More explicitly, we consider the geodesic equations of a massive particle moving around the black hole. Since the spacetime is stationary and axisymmetric, there are two corresponding constants of motion for geodesic equations, namely, the conserved energy $E$ and the azimuthal angular momentum $L_z$. Using these constants of motion, the $t$ and $\varphi$ components of the four-velocity of a moving particle can be expressed as  
\begin{align}
\dot{t}=\frac{Eg_{\varphi\varphi}+L_zg_{t\varphi}}{g_{t\varphi}^2-g_{tt}g_{\varphi\varphi}}\,,\label{dott}\\
\dot{\varphi}=-\frac{Eg_{t\varphi}+L_zg_{tt}}{g_{t\varphi}^2-g_{tt}g_{\varphi\varphi}}\,,
\end{align}
where the dot denotes the derivative with respect to the proper time of the particle. 

To obtain the expressions of the other two components, i.e., $\dot{r}$ and $\dot{y}$, we take the advantage of the separability of the geodesic equations. Due to the separability, the geodesic equations can be obtained from the Hamilton-Jacobi method. The geodesic equations of the Kerr-like metric in their first-order form have been obtained in Refs.~\cite{Chen:2020aix}. Although the author only considered the null geodesic, the generalization to time-like geodesic is straightforward. For the Kerr-like metric considered here, the $r$ and $y$ components of the four-velocity can be expressed as follows:
\begin{align}
\left[r^2+a^2y^2+\tilde\epsilon(y)\right]^2\dot{r}^2&=\mathcal{R}(r)\,,\label{Rdot}\\
\left[r^2+a^2y+\tilde\epsilon(y)\right]^2\dot{y}^2&=\mathcal{Y}(y)\,,\label{Ydot}
\end{align}
where
\begin{align}
\mathcal{R}(r)&=\left[E\left(r^2+a^2\right)-aL_z\right]^2-\left(\mathcal{K}+r^2\right)\Delta-\left(L_z-aE\right)^2\Delta\,,\\
\mathcal{Y}(y)&=\left[\mathcal{K}+\left(L_z-aE\right)^2-a^2y^2-\tilde\epsilon(y)\left(1-E^2\right)\right]\left(1-y^2\right)-\left[a\left(1-y^2\right)E-L_z\right]^2\,,
\end{align}
and $\mathcal{K}$ is a separation constant. The separation constant $\mathcal{K}$ plays the same role as the Carter constant in the original Kerr metric, therefore, we will still name it Carter constant throughout the paper. From now on, we will consider $n=1$, namely, $\tilde\epsilon(y)=\epsilon May$ in the rest of the paper. This choice is motivated not only by its simplicity, but also by the fact that in the models of effective field theory \cite{Cardoso:2018ptl,Cano:2019ore}, the terms in the metric that break reflection symmetry, at the leading order, usually take the forms proportional to $ay$. For this choice of the deviation function, the above equations \eqref{Rdot} and \eqref{Ydot} are invariant under the transformation $(\mathcal{K},\epsilon,y)\rightarrow(\mathcal{K},-\epsilon,-y)$.

\subsection{Circular orbits}
In general, circular motions still exist in this Kerr-like spacetime. However, due to the lack of $\mathbb{Z}_2$  symmetry, the circular motions are not located on the ``standard" equatorial plane $y=0$. On the contrary, each of them is confined on their own plane parallel to the standard equatorial plane, and the center of each orbit is on the axis of rotation of the black hole. Because each orbit has a constant radius, say, $r_c$, it can be labeled by their polar angle $\theta_c$ such that the distance between the standard equatorial plane and the plane in which the orbit resides is given by $h_c=r_c\cos\theta_c\equiv r_cy_c$. Furthermore, the $y_c$ of each orbit depends on the associated radius of the orbit $r_c$. To find the relation between $r_c$ and $y_c$ among the orbits, we start with the radial geodesic equation \eqref{Rdot}. Circular motions are characterized by the conditions $\mathcal{R}=d\mathcal{R}/dr=0$. Because the radial equation in our case is completely identical with that of the Kerr spacetime, the energy and azimuthal angular momentum for general spherical orbits can be parameterized by their radius $r$ and the Carter constant of the particle. The results have been derived in Ref.~\cite{Rana:2019bsn} (see Refs.~\cite{Hughes:1999bq,Hughes:2001jr,Fayos:2007ks} for different parameterizations) and they read \cite{Teo:2020sey}
\begin{align}
E&=\frac{r^3\left(r-2M\right)-a\left(a\mathcal{K}\mp\sqrt{\Gamma}\right)}{r^2\sqrt{r^3\left(r-3M\right)-2a\left(a\mathcal{K}\mp\sqrt{\Gamma}\right)}}\,,\label{Eradial}\\
L_z&=-\frac{2Mar^3+\left(r^2+a^2\right)\left(a\mathcal{K}\mp\sqrt{\Gamma}\right)}{r^2\sqrt{r^3\left(r-3M\right)-2a\left(a\mathcal{K}\mp\sqrt{\Gamma}\right)}}\,,\label{Lzradial}
\end{align}
where
\begin{equation}
\Gamma\equiv Mr^5-\mathcal{K}\left(r-3M\right)r^3+a^2\mathcal{K}^2\,.
\end{equation}

As we have just mentioned, a circular motion with radius $r=r_c$ is associated with a certain polar angle $y=y_c$. The value of $y_c$ can be obtained by combining Eqs.~\eqref{Eradial} and \eqref{Lzradial} with the equations $\mathcal{Y}=d\mathcal{Y}/dy=0$. Then, given the values of $a$ and $\epsilon$, we get a set of two coupled equations which are functions of $y$, $r$, and $\mathcal{K}$. Therefore, we can obtain the relation between $r_c$ and $y_c$ numerically. The results are shown in FIG.~\ref{fig.ycrc}. In the left panel, one can see that each circular motion is associated with a polar angle $y=y_c\ne0$. The value of $y_c$ depends on the radius of the circular motion $r_c$, and it approaches zero when $r_c$ is large, a manifestation of asymptotic flatness of the spacetime. For given values of $\epsilon$ and $a$, the accretion disk is composed of the collection of all circular orbits around the black hole. Most interestingly, due to the lack of reflection symmetry in the Kerr-like spacetime, instead of a flat disk confined on the equatorial plane, the accretion disk is a curved surface with curvature getting larger near the inner edge of the disk (see the right panel of FIG.~\ref{fig.ycrc}).

\begin{figure}[t!]
\centering
\includegraphics[angle =0,scale=0.5]{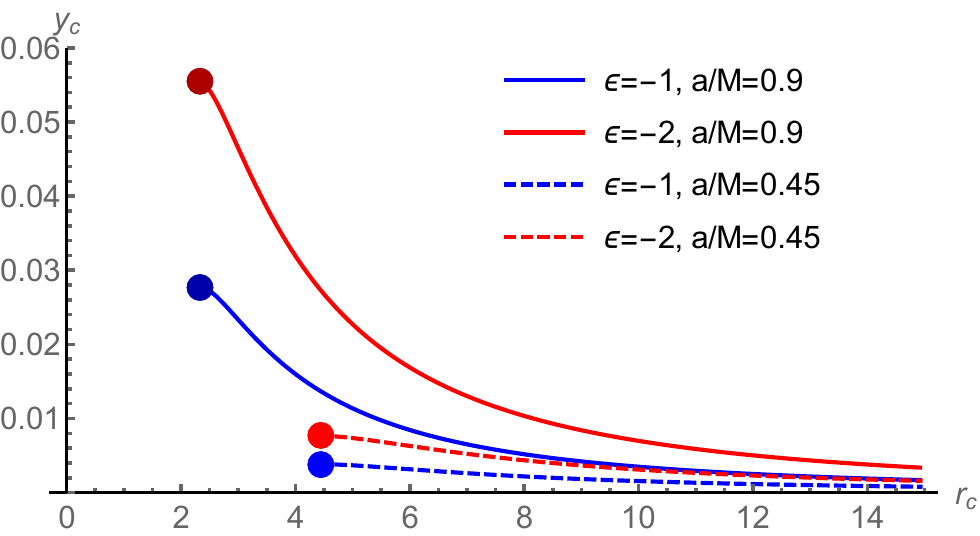}
\includegraphics[angle =0,scale=0.4]{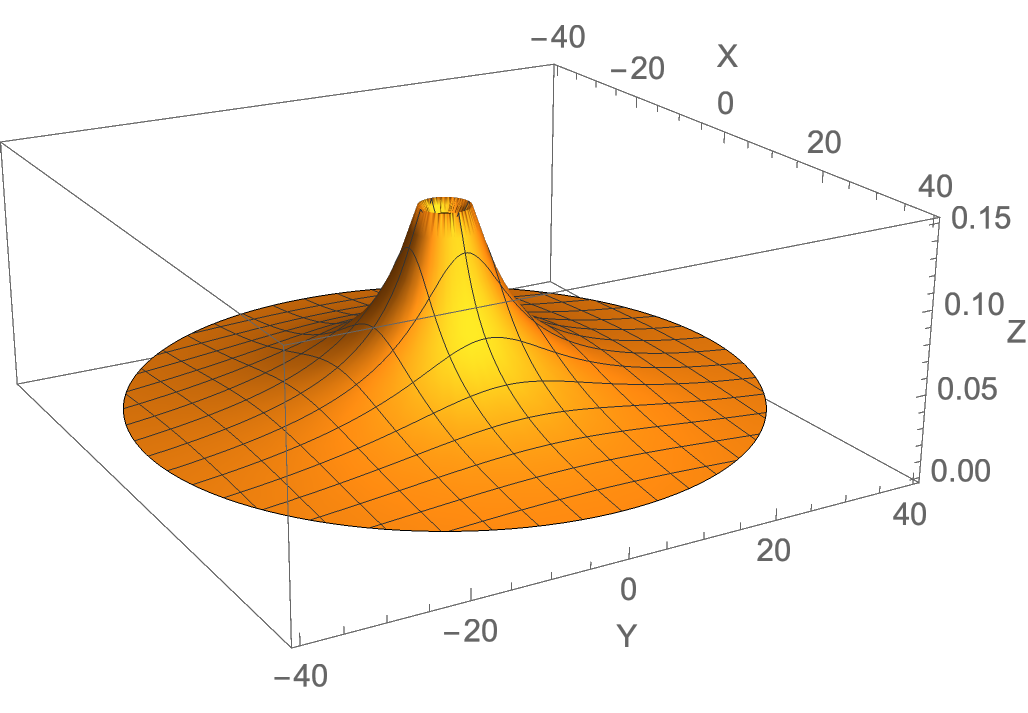}
\caption{(Left): In the presence of the deviation function $\tilde\epsilon(y)$, the circular motion of a massive particle moving around the black hole is associated with a polar angle $\theta_c$ which is not $\pi/2$. The value of $\theta_c$ depends on the radius of the circular orbit $r_c$. In this figure, we choose $\tilde\epsilon(y)=\epsilon May$, namely, $n=1$ in Eq.~\eqref{deviationepsilon}. All curves approach the standard equatorial plane $y_c\rightarrow 0$ when $r_c$ gets large. Also, each curve terminates at the ISCOs where the radial instability starts to occur (see the circular points on the left of each curve). It should be mentioned that the curves are symmetric with respect to the changes $(\epsilon,y_c)\rightarrow(-\epsilon,-y_c)$. (Right): The 3-dimensional plot of the accretion disk in the Cartesian coordinates $(X,Y,Z)=(r\sin\theta\cos\varphi,\,r\sin\theta\sin\varphi,\,r\cos\theta)$. We choose $\epsilon=-2$ and $a/M=0.9$. The disk is composed of the collection of all circular orbits. Notice that the coordinate along the $Z$-axis is amplified. The length in these figures is rescaled with respect to $M$.}
\label{fig.ycrc}
\end{figure}

\subsection{ISCO}
In the previous subsection, we have discussed the general property of the circular orbits of massive particles around the Kerr-like black hole. We find that in general, the circular orbits are not located on the standard equatorial plane. Also, the accretion disk composed of these circular orbits is actually a curved surface, as shown in FIG.~\ref{fig.ycrc}.

The disk property is highly related to the locations of the inner edge of the disk. The most simplest and common assumption is that the inner edge of the disk is determined by the ISCO of the black hole. The ISCO is the location where the circular orbit is marginally stable in the radial direction. Outside the ISCO, the particles can maintain stable quasi-circular motions and gradually spiral inward. On the other hand, after the particles cross the ISCO, the circular motions are radially unstable and they start to plunge into the black hole.

To identify the ISCO, we consider small perturbations $\delta r$ and $\delta\theta$ away from a circular orbit and obtain the following equations of motion:
\begin{align}
\frac{d^2\delta r}{dt^2}+\Omega_r^2\delta r=0\,,\qquad \frac{d^2\delta\theta}{dt^2}+\Omega_\theta^2\delta\theta=0\,,
\end{align}
where $\Omega_r$ and $\Omega_\theta$ are two epicyclic frequencies. They can be obtained numerically once $y_c$ and $r_c$ are solved using the method introduced in the previous subsection. Then, we identify the radial locations where $\Omega_r=0$ for a given set of $a$ and $\epsilon$. In FIG.~\ref{fig.iscover}, we focus on the prograde orbits and show the contours of constant ISCO (black solid curves) in the parameter space of $(a,|\epsilon|)$. It can be seen that for a fixed value of spin $a/M>0$, a non-zero $|\epsilon|$ shrinks the radius of ISCO, rendering the moving particles more likely to get close to the central black hole. The retrograde ISCOs are stretched by the deviation parameter, but the effects are largely suppressed hence not shown here. In this figure, we also calculate the polar angle $y_c$ on the ISCO. The contour of constant $|y_c|$ is shown by the dashed red curves, from left to right with $|y_c|=0.1$, $0.2$, and $0.3$. In addition, the purple dashed curve, which starts from the point $(a/M,|\epsilon|)\approx(0.943,0)$, stands for the parameter space in which the ISCO is located on the ergosurface. Furthermore, as we have mentioned in the previous section, for a given spin $a$ and if $|\epsilon|>\epsilon_{\textrm{sin}}$, naked singularities would start to appear from one of the poles. The magenta curve shows this critical value $\epsilon_{\textrm{sin}}$, i.e., Eq.~\eqref{epsiloncritical}, in the parameter space. Above this magenta curve, the naked singularity always appears somewhere outside the event horizon. For example, the thick blue curve refers to the parameter space in which the ISCO is located on the naked singularity. Therefore, for the whole accretion disk to be well-defined, we shall exclude the parameter space on the right of this thick blue curve.  

Finally, our numerical results show that there is no vertical instability for circular orbits around the Kerr-like black hole. In fact, for a fixed spin value $a$, we find that the deviation parameter $|\epsilon|$ would increase $\Omega_\theta^2$, which is already always positive for Kerr spacetime.

\begin{figure}[t!]
\centering
\includegraphics[angle =0,scale=0.7]{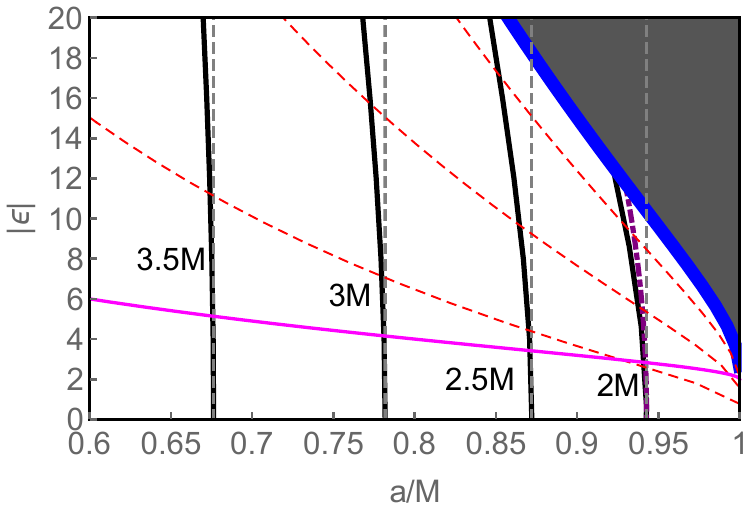}
\caption{The contours of constant ISCO (solid black) for the circular motion in the ($a,|\epsilon|$) parameter space. On the ISCO, we calculate constant $|y_c|$ and show them with dashed red curves ($|y_c|=0.1$, $0.2$, and $0.3$ from left to right). The vertical gray dashed lines show the constant ISCO for Kerr black holes. The magenta curve indicates the parameter space where the naked singularity starts to appear at the pole. The blue solid (purple dashed) curve indicates the parameter space where ISCO is located on the naked singularity (ergosurface). The region on the right of the blue curve is excluded. }
\label{fig.iscover}
\end{figure}

\subsection{Arbitrary deviation function $\tilde\epsilon(y)$}\label{sec.arbe}

Before discussing the astrophysical implications in the next section, let us move one step backward by relaxing the assumption on the deviation function $\tilde\epsilon(y)$ for a moment. We would like to emphasize that although most of the analysis conducted in this paper is based on a specific choice of the deviation function \eqref{deviationepsilon}, the fact that the accretion disk becomes curved and the prograde ISCOs are shrunk is actually generic regardless of the choice of the deviation function, at least when the reflection asymmetry is not large. We will show this by first assuming $\tilde\epsilon(0)=0$ and expanding the equations that determine ISCO radii to second-order in $y$ and $\mathcal{K}$, which are supposed to be zero for circular orbits around a black hole with $\mathbb{Z}_2$ symmetry. Then, by assuming that $\tilde\epsilon'(0)$ and $\tilde\epsilon''(0)$, where the primes denote derivatives with respect to $y$, are of the same order of $y$, we obtain (by setting $M=1$ for simplicity)
\begin{align}
&\mathcal{K}\approx\frac{y\tilde\epsilon'(0)\left(r^2-4r+4a\sqrt{r}-a^2\right)+y^2\left(r^4-4ar^{5/2}+3r^2a^2\right)}{2ar^{3/2}+r^3-3r^2}\,,\\
&\tilde\epsilon'(0)\approx\frac{2\left(r^4-4ar^{5/2}+3r^2a^2\right)y}{4r-r^2-4a\sqrt{r}+a^2}\,,
\end{align}
meaning that $\mathcal{O}(\mathcal{K})\sim\mathcal{O}(y^2)\sim\mathcal{O}(y\tilde\epsilon'(0))$. Using these equations, one can obtain the following approximated formula to determine the ISCO radii:
\begin{equation}
\frac{-6r+r^2+8a\sqrt{r}-3a^2}{r^2\left(r^{3/2}+a\right)^2}+\frac{y^2}{r^4\left(r^2-4r+4\sqrt{r}a-a^2\right)\left(r^{3/2}+a\right)^3}\mathcal{W}(r,a)=0\,,\label{generaliscoeq1st}
\end{equation}
where 
\begin{align}
\mathcal{W}(r,a)=&\,48 r^{11/2} - 32 r^{13/2} + 4 r^{15/2} - 216 ar^4  + 
  34 ar^5  + 23 ar^6  + 192 a^2r^{5/2}  \nonumber\\&+ 
  472 a^2r^{7/2}  - 148 a^2r^{9/2} - 
  6 a^2r^{11/2} - 504 a^3r^2  - 440 a^3r^3 + 
  123 a^3r^4  + 504 a^4r^{3/2} \nonumber\\ &+ 
  228 a^4r^{5/2}  - 36 a^4r^{7/2}  - 234 a^5r  - 
  63 a^5r^2 + 48 a^6\sqrt{r}  + 6 a^6r^{3/2}  - 
  3 a^7\,.
\end{align}
The ISCO of a Kerr black hole can be calculated by solving Eq.~\eqref{generaliscoeq1st} with $y=0$. In this case, all circular orbits are confined on the equatorial plane. On the other hand, for deviation functions that break the reflection symmetry, the circular orbits are not on the equatorial plane, i.e., $y\ne0$, and therefore, the ISCO would differ from their Kerr counterparts. 

In table~\ref{summaryiscovalues}, we calculate the radii of some prograde ISCOs by choosing black holes spins $a/M=0.8$ and $0.9$, and different values of $y$ on which the orbits reside. The ISCO radii are calculated in two ways: the approximated formula \eqref{generaliscoeq1st}, in which the deviation function remains quite arbitrary, and our specific model whose deviation function is given by Eq.~\eqref{deviationepsilon} with $n=1$. We also calculate the relative difference between the values obtained from these two approaches:
\begin{equation}
\textrm{Rel. difference}= \frac{r_{ISCO}[\textrm{App. by Eq.~\eqref{generaliscoeq1st}}]}{r_{ISCO}[\textrm{Model $n=1$}]}-1\,.
\end{equation}
One clearly sees that the approximated formula \eqref{generaliscoeq1st} does a rather good job in capturing the shrinking of prograde ISCOs in the presence of reflection asymmetry. It suggests that, at least in the regime where the asymmetry is small, the shrinking of prograde ISCOs is independent of the choice of the deviation function $\tilde\epsilon(y)$. Finally, the relative difference between the two approaches increases when $|y|$ increases. This is due to the gradual failure of the approximated formula \eqref{generaliscoeq1st} when going beyond $|y|\ll1$ approximations, and hence is completely expected.

\begin{table*}
 \begin{center}
  \begin{tabular}{c|c|c|c||c|c|c||}
  \cline{2-7}
  &\multicolumn{3}{|c||}{$a/M=0.8$}&\multicolumn{3}{|c|}{$a/M=0.9$}\\ 
 \cline{2-7}
  &\multicolumn{1}{|c|}{App. by Eq.~\eqref{generaliscoeq1st}}&\multicolumn{1}{|c|}{Model $n=1$}&\multicolumn{1}{|c||}{Rel. difference}&\multicolumn{1}{|c|}{App. by Eq.~\eqref{generaliscoeq1st}}&\multicolumn{1}{|c|}{Model $n=1$}&\multicolumn{1}{|c||}{Rel. difference}\\ 
   \hline\hline
   \multicolumn{1}{||c||}{$|y|=0$ (Kerr)}  & $2.90664M$ & $2.90664M$ & $0\%$ & $2.32088M$  & $2.32088M$ & $0\%$  \\ \hline
   \multicolumn{1}{||c||}{$|y|=0.1$}  & $2.89715M$ & $2.89634M$ & $0.028\%$  & $2.31158M$ & $2.31092M$ & $0.029\%$  \\ \hline
    \multicolumn{1}{||c||}{$|y|=0.2$}  & $2.87336M$ & $2.86094M$ & $0.43\%$ & $2.28754M$ & $2.27725M$ & $0.45\%$ \\ \hline
     \multicolumn{1}{||c||}{$|y|=0.3$}  & $2.84474M$  & $2.78341M$ & $2.2\%$ & $2.25709M$  & $2.20632M$ &$2.3\%$ \\ \hline
     \end{tabular}
  \caption{The values of prograde ISCO radii obtained from the approximated formula \eqref{generaliscoeq1st}, in which the deviation function remains quite arbitrary, and those obtained from our specific model with $n=1$. The relative difference between the values obtained from the two approaches is calculated.}
    \label{summaryiscovalues}
 \end{center}
\end{table*}

\section{Astrophysical implications}\label{sec.astr}
After discussing the general properties of the circular orbits as well as the accretion disk around the Kerr-like black hole, one would naturally ask how the equatorial reflection asymmetry could be possibly detected using astrophysical observations. In Ref.~\cite{Chen:2020aix}, the spherical photon orbits and the shadow cast by the Kerr-like black holes have been discussed. The spherical photon orbits of Kerr-like black holes can be defined on constant-$r$ surfaces in the Boyer-Lindquist coordinates. The photon trajectories on these surfaces are asymmetric with respect to the equatorial plane (see Figure 5 in Ref.~\cite{Chen:2020aix}). In addition, it has been demonstrated that the contour of the shadow cast by the Kerr-like black hole can be a useful tool to probe the parameter $\epsilon$. More explicitly, the deviation parameter $\epsilon$ is sensitive to the apparent size of the shadow contour, but it does not significantly affect the distortion in shape of the contour. Because the shape distortion of the shadow contour is sensitive to the black hole spin and the inclination angle of the observer, the shadow size measurement, combined with other independent measurements of mass and distance, can be used to constrain $\epsilon$, without worrying about its degeneracy with the black hole spin $a$. However, due to the small angular sizes of the black hole shadows, the number of shadows currently observable is extremely limited even for state-of-the-art very-long-baseline-interferometer technology such as the Event Horizon Telescope \cite{EventHorizonTelescope:2019dse}. In addition, the capability of constraining parameters beyond Kerr metric using the shadow size measurements \cite{EventHorizonTelescope:2020qrl,Brahma:2020eos,Shaikh:2021yux,EventHorizonTelescope:2021dqv,Zakharov:2021gbg} is still far from being conclusive \cite{Gralla:2020pra,Volkel:2020xlc,Glampedakis:2021oie,Nampalliwar:2021oqr} (see Ref.~\cite{Lima:2021las} for some issues about degeneracy). Therefore, in this section, we will consider the astrophysical imprints of the equatorial reflection asymmetry on the accretion disk and its associated measurements.

\subsection{Radiative efficiency}
During the mass accreting process, particles orbiting the central black hole convert their rest-mass energy into radiation. In the standard thin-disk model \cite{Page:1974he,Thorne:1974ve}, the energy conversion can be very efficient. One can define the radiative efficiency $\eta$ to be the ratio between the rate at which the photon energy escapes from the disk to infinity, and the transfer rate of mass energy, both calculated at infinity. The radiative efficiency calculated in this way ranges from $\sim5.7\%$ for Schwarzschild black holes to $\sim40\%$ for maximally spinning black holes. For nearly extremal black holes with $a\approx M$, the radiative efficiency grows rapidly due to the shrinking of the ISCO, allowing more efficient energy extractions from the moving particles.

Assuming that all photons emitted from the disk surface can escape to spatial infinity, and neglecting the back reaction of radiation on the black hole, the radiative efficiency can be expressed as \cite{Thorne:1974ve}
\begin{equation}
\eta=1-E|_{\textrm{ISCO}}\,.\label{etadefinition}
\end{equation}
We show the radiative efficiency $\eta$ with respect to the deviation parameter $\epsilon$ and the black hole spin $a$ in FIG.~\ref{fig.eta}. We recover the canonical result that, for Kerr black holes, the radiative efficiency increases with the black hole spin, especially when the spin is near extremal. For Kerr-like black holes, the deviation parameter $\epsilon$ can further increase the radiative efficiency, and the effect is more pronounced for larger spin values. For instance, the efficiency could grow from $\eta\sim0.32$ for $|\epsilon|=0$ to $\eta\sim0.39$ for $|\epsilon|=3$ for $a/M=0.9982$, i.e., a $20\%$ increase. Recall that for a fixed spin, a non-vanishing $\epsilon$ would shrink the size of the prograde ISCO. Therefore, particles undergoing prograde circular motions around the Kerr-like black hole can get even closer to the central hole and hence more energies can be extracted, as compared with the cases in Kerr spacetime. 

The different location of the ISCO and resulting radiative efficiency could have impacts on the black hole spin measurements inferred from the continuum fitting method. Assuming the Kerr metric and the standard thin-disk model, the ISCO could be fitted from the continuum emission from the accretion disk \cite{Shafee:2006,McClintock:2006}. Given the curved accretion disk and the smaller prograde ISCO of Kerr-like black holes, it may have non-trivial effects on the spin constraints. However, detailed models of curved accretion disks in the absence of reflection symmetry have not yet existed, and we will delay the development of such a model to future studies.

It should be emphasized that the radiative efficiency given in Eq.~\eqref{etadefinition} is strictly an upper limit due to the negligence of the photon capturing effects \cite{Thorne:1974ve}. The calculation of the radiative efficiency taking into account the photon captures requires a reliable thin-disk model for a ``curved" disk, whose formulation is beyond the scope of this paper. However, the qualitative result that the radiative efficiency is enhanced by $\epsilon$ should still hold in general because the prograde ISCO being shrunk by $\epsilon$ is a generic property of the model.

\begin{figure}[t!]
\centering
\includegraphics[angle =0,scale=0.39]{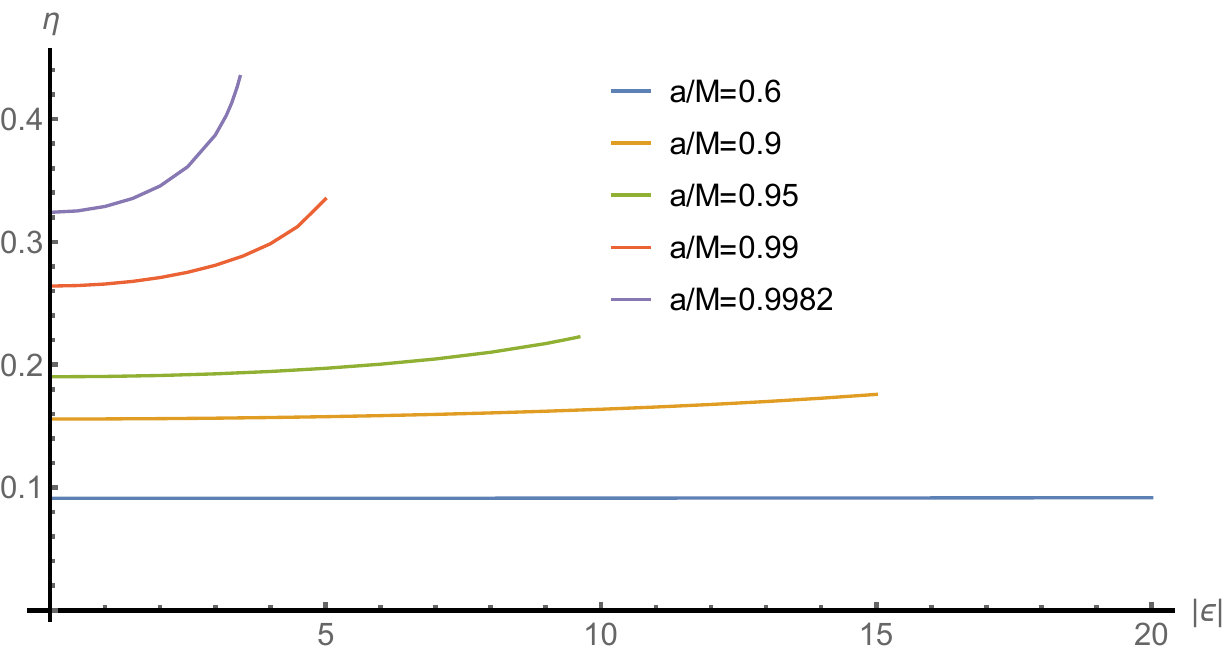}
\includegraphics[angle =0,scale=0.39]{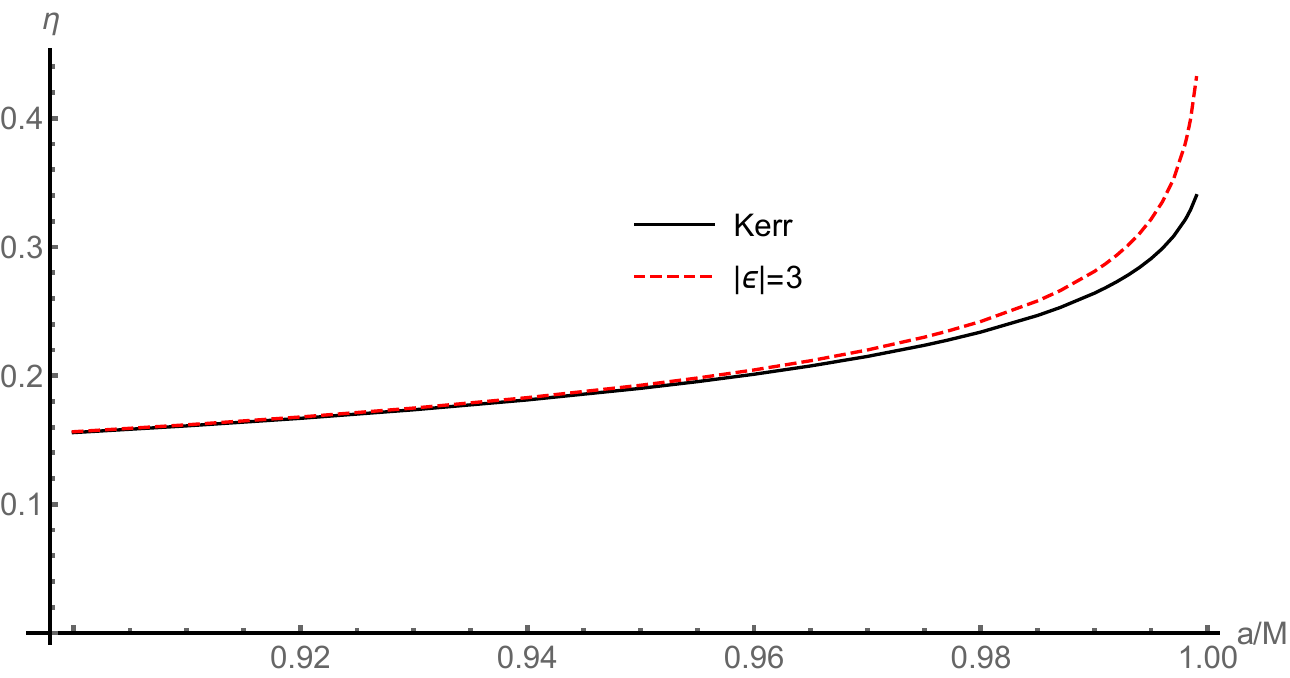}
\caption{(Left): The radiative efficiency $\eta$ as a function of the deviation parameter $|\epsilon|$. Different curves represent different values of the black hole spin, as indicated by the legend. Each curve terminates at a certain $|\epsilon|$ when the ISCO coincides with the naked singularity. (Right): The radiative efficiency $\eta$ for Kerr (black) and Kerr-like black holes (red dashed), as a function of the black hole spin. The further enhancement of $\eta$ for Kerr-like black holes at large spins is clearly seen.}
\label{fig.eta}
\end{figure}

\subsection{Gravitational redshift and $g$-factor}

Another important effect to quantify is the redshift of photons emitted from the accretion disk. In general, for observers in the locally nonrotating frame, i.e., zero angular momentum observers (ZAMO), observed frequencies of photons coming from the disk would be shifted with respect to the frequencies locally measured at the emission. The relative motion of particles on the disk generates Doppler shifting and the relativistic Doppler beaming. In addition, there are significant gravitational redshifts induced by the strong gravitational fields near the black hole. Therefore, the redshift highly depends on the disk property, including the location of the emission, the inclination angle, and the parameters associated with the black hole. The overall redshift measurements can be used to estimate the spin of the black hole \cite{Reynolds:2019uxi} and even to test the Kerr hypothesis. In the following, we investigate the influence on gravitational redshifts due to spacetime distortions near Kerr-like black holes.

One can define a quantity, so-called $g$-factor, which is related to the gravitational redshift, $z$, by \cite{Dabrowski:2000qv,Mueller:2003fr,Mueller:2006dn,Bambi:2017khi}:
\begin{equation}
g=\frac{1}{1+z}\,.\label{gz}
\end{equation}
To derive the general expression of the $g$-factor, we consider a general axisymmetric and stationary spacetime, whose metric can be recast into the following expression:
\begin{equation}
ds^2=-\alpha^2dt^2+\tilde\omega^2\left(d\varphi-\omega dt\right)^2+\frac{\rho^2}{\Delta}dr^2+\frac{\rho^2}{1-y^2}dy^2\,,\label{metricalphaex}
\end{equation}
where the metric functions $\alpha$, $\tilde\omega$, $\omega$, $\rho$, and $\Delta$ are given by
\begin{align}
\alpha&=\sqrt{\frac{g_{t\varphi}^2-g_{tt}g_{\varphi\varphi}}{g_{\varphi\varphi}}}\,,\quad\tilde\omega=\sqrt{g_{\varphi\varphi}}\,,\quad\omega=-\frac{g_{t\varphi}}{g_{\varphi\varphi}}\,,\nonumber\\
\rho^2&=g_{yy}\left(1-y^2\right)\,,\qquad\Delta=\frac{g_{yy}\left(1-y^2\right)}{g_{rr}}\,.
\label{alphametric}
\end{align}

The redshift factor is measured by an observer in the locally nonrotating frame (ZAMO). One can define the tetrad field $e_{0(a)}^\mu$ of that frame such that \cite{Bini:2001kx}
\begin{equation}
g^{\mu\nu}=e_{0(a)}^\mu e_{0(b)}^\nu \eta^{(a)(b)}\,,\label{tetraddef}
\end{equation}
where $\eta^{(a)(b)}=\textrm{diag}(-1,1,1,1)$ is the inverse of the Minkowskian metric. The subscript 0 in the tetrad denotes the locally nonrotating frame. Inserting the metric \eqref{metricalphaex} into Eq.~\eqref{tetraddef}, the tetrad field can be expressed as
\begin{align}
e^{\mu}_{0(t)}&=\left(1/\alpha,\quad0,\quad0,\quad\omega/\alpha\right)\,,\nonumber\\
e^{\mu}_{0(r)}&=\left(0,\quad \sqrt{\Delta}/\rho,\quad 0,\quad0\right)\,,\nonumber\\
e^{\mu}_{0(y)}&=\left(0,\quad 0,\quad \sqrt{1-y^2}/\rho,\quad0\right)\,,\nonumber\\
e^{\mu}_{0(\varphi)}&=\left(0,\quad 0,\quad 0,\quad 1/\tilde\omega\right)\,.\label{tetradbasiszamo}
\end{align}

We assume that the emission is from the accreting particles undergoing circular orbits around the black hole. The ZAMO frame can then be boosted into the locally rest frame of the accreting particles $e_{(a)}^\mu$:
\begin{align}
e^{\mu}_{0(t)}&=\gamma\left(e_{(t)}^\mu-v^{(\varphi)}e_{(\varphi)}^\mu\right)\,,\nonumber\\
e^{\mu}_{0(r)}&=e^{\mu}_{(r)}\,,\nonumber\\
e^{\mu}_{0(y)}&=e^{\mu}_{(y)}\,,\nonumber\\
e^{\mu}_{0(\varphi)}&=\gamma\left(-v^{(\varphi)}e_{(t)}^\mu+e_{(\varphi)}^\mu\right)\,,\label{tetradbasisptl}
\end{align}
where $v^{(\varphi)}$ is the azimuthal velocity of disk particles relative to the ZAMO frame, and $\gamma=\sqrt{1-v^{(\varphi)2}}$ is the Lorentz factor. Note that $e_{(t)}^\mu=u_e^\mu=\dot{t}(1,0,0,\Omega)$ is the 4-velocity of the particles undergoing circular motions, where $\Omega=\dot\varphi/\dot{t}$ is the angular frequency of the accreting particles. Since the emission is coming from these particles, we include a subscript ``e".

By eliminating $e_{(\varphi)}^\mu$ from Eqs.~\eqref{tetradbasiszamo} and \eqref{tetradbasisptl}, one gets
\begin{align}
e_{0(t)}^\mu+v^{(\varphi)}e_{0(\varphi)}^\mu&=\frac{1}{\gamma}e_{(t)}^\mu=\frac{1}{\gamma}\dot{t}\left(1,0,0,\Omega\right)\nonumber\\
&=\frac{1}{\alpha}\left(1,0,0,\omega+\alpha v^{(\varphi)}/\tilde\omega\right)\,.\label{26e0}
\end{align}
Then, one can also obtain the following formulas
\begin{equation}
\dot{t}=\frac{\gamma}{\alpha}\,,\qquad v^{(\varphi)}=\frac{\tilde\omega}{\alpha}\left(\Omega-\omega\right)\,.
\end{equation}

The redshift factor $g$ is defined by the ratio between the energy measured by an observer with 4-velocity $u_o^\mu=(1,0,0,0)$ and that measured by the local frame of the accreting particles:
\begin{equation}
g\equiv\frac{-u_o^\mu k_\mu}{-u_e^\nu k_\nu}=\frac{E_o}{E_e}\,.
\end{equation}
In the above expression, we define $E_e\equiv-u_e^\nu k_\nu$ where $k_\mu=(-E_o,0,0,L_{zo})$ is the 4-momentum of photons, in which $E_o$ and $L_{zo}$ are conserved energy and azimuthal angular momentum of photons. Then, using Eq.~\eqref{26e0}, we have
\begin{equation}
-\frac{E_e}{\gamma}=\frac{1}{\alpha}\left(-E_o+L_{zo}\Omega\right)\,.
\end{equation}
Finally, the redshift $g$-factor can be written as
\begin{equation}
g=\frac{\alpha}{\gamma\left(1-\lambda_o\Omega\right)}=\frac{1}{\dot{t}\left(1-\lambda_o\Omega\right)}\,,
\end{equation}
where $\lambda_o\equiv L_{zo}/E_{o}$ is the specific angular momentum of the emitted photons and $\dot{t}$ is defined by Eq.~\eqref{dott}.

\begin{figure}[t!]
\centering
\includegraphics[angle =0,scale=0.7]{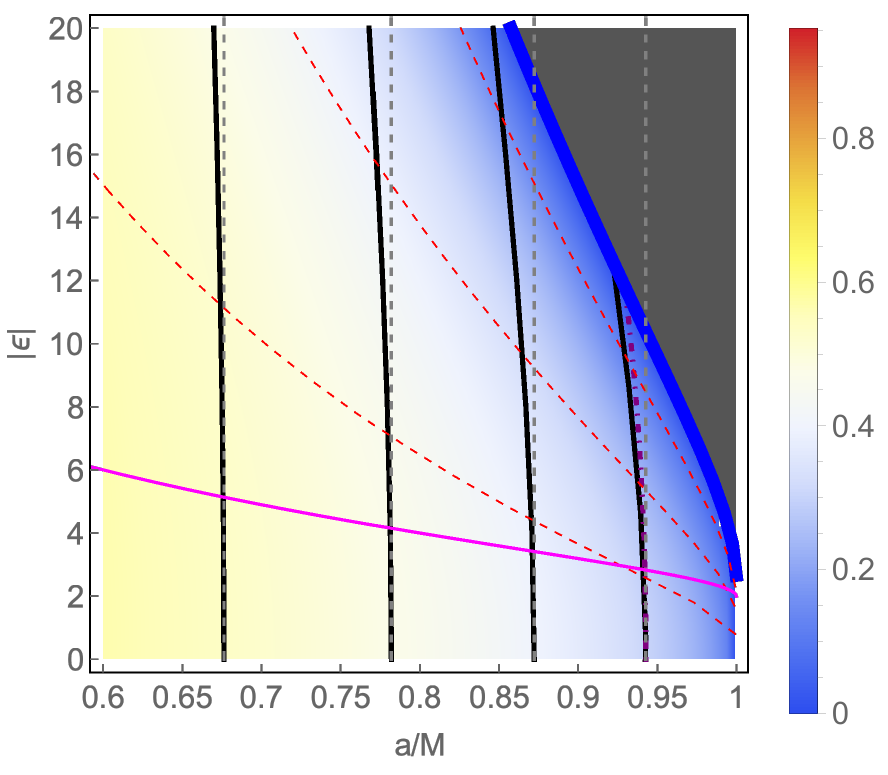}
\caption{This density contour plot shows the $g$-factor measured by a face-on observer ($\lambda_o$=0), depicted on the ($a,|\epsilon|$) plane. The emission is assumed to be from the accreting particles on the ISCO. The solid black and the dashed gray curves represent the contours of constant ISCO for Kerr-like and the Kerr black holes, respectively. On the ISCO, the constant-$|y_c|$ contours are shown by the dashed red curves. The magenta curve shows the critical value given by Eq.~\eqref{epsiloncritical}. The blue solid (purple dashed) curve indicates the parameter space where ISCO is located on the naked singularity (ergosurface).}
\label{fig.gfactor}
\end{figure}

In FIG.~\ref{fig.gfactor}, we consider an observer plane parallel to the plane of circular motions, namely, a face-on observer. The redshift $g$-factor is calculated by assuming that the photons are emitted from the accreting particles orbiting at the ISCO. For photons emitted from the accretion disk and reaching the face-on observer, their specific angular momentum $\lambda_o$ is zero. The meaning of the dashed red, purple, and gray curves, as well as the solid blue, black, and magenta curves can be read off from the caption of FIG.~\ref{fig.iscover}. One can immediately see from FIG.~\ref{fig.gfactor} that the $g$-factor vanishes both at the horizon and the naked singularity. According to the definition of the $g$-factor in Eq.~\eqref{gz}, a vanishing $g$-factor indicates that the gravitational redshift $z$ is infinite.

As can be seen from FIG.~\ref{fig.gfactor}, the prograde ISCO of a black hole with a high spin but a small $|\epsilon|$ can give the same $g$-factor as that of a black hole with a smaller spin but larger $|\epsilon|$. For example, the $g$-factor emitted from the ISCO of a Kerr black hole with $a/M=0.99$ is $g\approx0.164$. The same $g$-factor can be obtained from the ISCO of a Kerr-like black hole with $a/M=0.85$ and $|\epsilon|=17.37$. A direct consequence is that the black hole spin measurement based on the redshift $g$-factor, assuming the Kerr hypothesis, could overestimate the true spin value if the black hole is actually described by a Kerr-like metric with a large $|\epsilon|$. In the above example, the black hole spin could be overestimated by $\sim16.4\%$, a non-negligible effect if $|\epsilon|$ is large. Of course, for a large $|\epsilon|$, the spacetime would suffer from the existence of naked singularities somewhere. However, as long as one does not consider the parameter space inside the gray shaded region, the naked singularity is always inside ISCO and the whole accretion disk can be completely well-defined. In fact, significant effects on the radiative efficiency and the redshifts may still appear even if one strictly excludes any possible existence of naked singularity and focuses only on $|\epsilon|\le2$. For example, the radiative efficiency $\eta$ for a Kerr-like black hole with $a/M=0.9982$ and $|\epsilon|=2$ is enhanced compared with its Kerr counterpart with the same spin by $6.6\%$. In addition, the redshift $z$ calculated on the prograde ISCO for a Kerr-like black hole with $a/M=0.9982$ and $|\epsilon|=2$ is $z\sim14.83$, which is larger than its Kerr counterpart with the same spin ($z\sim10.2$) by $45\%$.

\section{Conclusions}\label{sec.conclusion}

In this paper, we adopt a theory-agnostic approach and consider a class of Kerr-like spacetime, which was recently proposed in Ref.~\cite{Chen:2020aix}, to model a black hole without equatorial reflection symmetry. 
After requiring that the spacetime reduces to Kerr metric at spatial infinity and that the metric satisfies the Solar System constraints, the Kerr-like metric only contains a deviation function $\tilde\epsilon(y)$, which can generically break the reflection symmetry of the spacetime. Rotating black hole solutions without $\mathbb{Z}_2$ symmetry naturally arise in effective theories containing parity-violating interactions, and the asymmetry is essentially induced by the spin of the black hole. The Kerr-like metric considered here is relatively simple in that its geodesic equations are designed to be completely separable. Therefore, the metric can be a good approximation of those complicated solutions in effective theories and can be very useful in studying the astrophysical implications of reflection asymmetry in a phenomenological manner.
  
We investigate the properties of accretion disk around the Kerr-like black hole. We start with the geodesic equations and focus on the circular orbits of massive particles in the spacetime. Due to the reflection asymmetry, the accretion disk, which is composed of all circular orbits collected from the ISCO to some large radius, is generically a curved surface, rather than a flat disk confined on the standard equatorial plane $y=0$. In fact, it has been shown in Ref.~\cite{Datta:2020axm} that in the absence of equatorial reflection symmetry, circular orbits confined on the equatorial plane do not exist. Their result is consistent with ours because in our model, the circular orbit with radius $r_c$ is in general associated with a polar angle $y_c\ne0$, namely, it is confined on a plane parallel to, but different from, the equatorial plane. The orbit with a large radius gradually approaches the equatorial plane, i.e., $y_c\rightarrow0$ when $r_c\rightarrow\infty$. This is a manifestation of asymptotic flatness of the spacetime (FIG.~\ref{fig.ycrc}). 

We explore the astrophysical implications of $\mathbb{Z}_2$ asymmetry on the accretion disk properties around a Kerr-like black hole. In particular, we find that, in a toy model with a specific choice of the deviation function, the parameter $\epsilon$ that controls $\mathbb{Z}_2$ asymmetry would shrink the radius of the prograde ISCO. Assuming that the inner edge of the disk is defined by the ISCO, this then means that the materials whirling on the prograde disk can get closer to the central black hole, as compared to the Kerr case. Therefore, the radiative efficiency of the accretion disk $\eta$, defined as the fraction of rest-mass energy converted into radiation, can be enhanced in the absence of reflection symmetry (FIG.~\ref{fig.eta}). Although the above analysis is done under a specific choice of the deviation function, in Sec.~\ref{sec.arbe}, we show that in the presence of only small reflection asymmetry, the shrinking of prograde ISCOs is actually quite generic in our model, independent of the choice of the deviation function. On the other hand, the asymmetry would stretch the retrograde ISCOs, but the effects are largely suppressed and could be unmeasurably small.

We finally investigate the gravitational redshift effect and compute the $g$-factor associated with the emission coming from the ISCO in the Kerr-like spacetime. The most important result is that the spin measurements based on the redshift $g$-factor observations should be analyzed with great care. Assuming the Kerr hypothesis, such measurements could overestimate the true spin value of the black hole if the black hole is actually the Kerr-like one with a large deviation parameter $|\epsilon|$ (at a level of $\sim16\%$ if $|\epsilon|\sim17$).

There are other important observables of the accretion disk that we have not explored in this study, such as the disk temperature and luminosity. In the cases in which the spacetime possesses equatorial reflection symmetry, one can adopt the thin-disk model and the calculations of these observables can be quite straightforward \cite{Page:1974he}. However, in the absence of reflection symmetry, the disk is a curved surface and one has to rebuild the corresponding curved thin-disk model. This is beyond the scope of the present paper. In addition, it will be interesting to investigate the detailed motions of particles after they cross the ISCO and enter the plunging phase. All the above explorations would give further insights into the fundamental differences between Kerr and Kerr-like black holes, and possibly their observable signatures. We will leave these interesting topics to future works.

\section*{Acknowledgement}

CYC is supported by the Institute of Physics of Academia Sinica. HYKY acknowledges support from Yushan Scholar Program of the Ministry of Education of Taiwan, and Ministry of Science and Technology of Taiwan (MOST 109-2112-M-007-037-MY3).


\begin{thebibliography}{99}


\bibitem{Johannsen:2010xs}
T.~Johannsen and D.~Psaltis,
Astrophys. J. \textbf{716}, 187-197 (2010).

\bibitem{Johannsen:2010ru}
T.~Johannsen and D.~Psaltis,
Astrophys. J. \textbf{718}, 446-454 (2010).

\bibitem{Johannsen:2010bi}
T.~Johannsen and D.~Psaltis,
Astrophys. J. \textbf{726}, 11 (2011).



\bibitem{Bambi:2011mj}
C.~Bambi,
Mod. Phys. Lett. A \textbf{26}, 2453-2468 (2011).

\bibitem{Johannsen:2012ng}
T.~Johannsen and D.~Psaltis,
Astrophys. J. \textbf{773}, 57 (2013).

\bibitem{Yagi:2016jml}
K.~Yagi and L.~C.~Stein,
Class. Quant. Grav. \textbf{33}, no.5, 054001 (2016).

\bibitem{Johannsen:2016uoh}
T.~Johannsen,
Class. Quant. Grav. \textbf{33}, no.12, 124001 (2016).

\bibitem{Barack:2018yly}
L.~Barack, V.~Cardoso, S.~Nissanke, T.~P.~Sotiriou, A.~Askar, C.~Belczynski, G.~Bertone, E.~Bon, D.~Blas and R.~Brito, \textit{et al.}
Class. Quant. Grav. \textbf{36}, no.14, 143001 (2019).

\bibitem{Bambi:2015kza}
C.~Bambi,
Rev. Mod. Phys. \textbf{89}, no.2, 025001 (2017).

\bibitem{Isi:2019aib}
M.~Isi, M.~Giesler, W.~M.~Farr, M.~A.~Scheel and S.~A.~Teukolsky,
Phys. Rev. Lett. \textbf{123}, no.11, 111102 (2019).



\bibitem{Shafee:2006}
R.~Shafee \textit{et al.},
Astrophys. J. Lett. \textbf{636}, L113 (2006).

\bibitem{McClintock:2006}
J.~E.~McClintock \textit{et al.},
Astrophys. J. \textbf{652}, 518 (2006).

\bibitem{Fabian:1989}
A.~C.~Fabian, M.~J.~Rees, L.~Stellar and N.~E.~White, 
Mon. Not. R. Astron. Soc. \textbf{238}, 729 (1989).

\bibitem{Brenneman:2006}
L.~W.~Brenneman and C.~S.~Reynolds, 
Astrophys. J. \textbf{652}, 1028 (2006).

\bibitem{Risaliti:2013}
G.~Risaliti \textit{et al.},
Nature \textbf{494}, 449 (2013).

\bibitem{Shakura:1973}
N.~I.~Shakura and R.~A.~Sunyaev,
Astron. Astrophys. \textbf{24}, 337 (1973).

\bibitem{Novikov:1973}
I.~D.~Novikov and K.~S.~Thorne,
Black Holes (Les Astres Occlus), 
Gordon and Breach Science Publishers, 343 (1973).


\bibitem{Pun:2008ae}
C.~S.~J.~Pun, Z.~Kov\'acs and T.~Harko,
Phys. Rev. D \textbf{78}, 024043 (2008).

\bibitem{Harko:2009rp}
T.~Harko, Z.~Kov\'acs and F.~S.~N.~Lobo,
Phys. Rev. D \textbf{80}, 044021 (2009).

\bibitem{Harko:2009kj}
T.~Harko, Z.~Kov\'acs and F.~S.~N.~Lobo,
Class. Quant. Grav. \textbf{27}, 105010 (2010).



\bibitem{Harko:2010ua}
T.~Harko, Z.~Kov\'acs and F.~S.~N.~Lobo,
Class. Quant. Grav. \textbf{28}, 165001 (2011).


\bibitem{Perez:2012bx}
D.~Perez, G.~E.~Romero and S.~E.~P.~Bergliaffa,
Astron. Astrophys. \textbf{551}, A4 (2013).

\bibitem{Soroushfar:2020kgb}
S.~Soroushfar and S.~Upadhyay,
Eur. Phys. J. Plus \textbf{135}, no.3, 338 (2020).



\bibitem{Liu:2020vkh}
C.~Liu, T.~Zhu and Q.~Wu,
Chin. Phys. C \textbf{45}, no.1, 015105 (2021).

\bibitem{Heydari-Fard:2021ljh}
M.~Heydari-Fard, M.~Heydari-Fard and H.~R.~Sepangi,
Eur. Phys. J. C \textbf{81}, no.5, 473 (2021).



\bibitem{Liu:2021yev}
C.~Liu, S.~Yang, Q.~Wu and T.~Zhu,
[arXiv:2107.04811 [gr-qc]].










\bibitem{Heydari-Fard:2020ugv}
M.~Heydari-Fard, M.~Heydari-Fard and H.~R.~Sepangi,
Eur. Phys. J. C \textbf{80}, no.4, 351 (2020).

\bibitem{Collodel:2021gxu}
L.~G.~Collodel, D.~D.~Doneva and S.~S.~Yazadjiev,
Astrophys. J. \textbf{910}, no.1, 52 (2021).





\bibitem{Pun:2008ua}
C.~S.~J.~Pun, Z.~Kov\'acs and T.~Harko,
Phys. Rev. D \textbf{78}, 084015 (2008).

\bibitem{Chen:2011wb}
S.~Chen and J.~Jing,
Phys. Lett. B \textbf{704}, 641-645 (2011).









\bibitem{Kovacs:2010xm}
Z.~Kov\'acs and T.~Harko,
Phys. Rev. D \textbf{82}, 124047 (2010).


\bibitem{Shahidi:2020bla}
S.~Shahidi, T.~Harko and Z.~Kov\'acs,
Eur. Phys. J. C \textbf{80}, no.2, 162 (2020).




\bibitem{Bambi:2011jq}
C.~Bambi and E.~Barausse,
Astrophys. J. \textbf{731}, 121 (2011).

\bibitem{Chen:2011rx}
S.~Chen and J.~Jing,
Phys. Lett. B \textbf{711}, 81-87 (2012).

\bibitem{Bambi:2012tg}
C.~Bambi,
Astrophys. J. \textbf{761}, 174 (2012).

\bibitem{Delgado:2021jxd}
J.~F.~M.~Delgado, C.~A.~R.~Herdeiro and E.~Radu,
[arXiv:2107.03404 [gr-qc]].






\bibitem{Glampedakis:2005cf}
K.~Glampedakis and S.~Babak,
Class. Quant. Grav. \textbf{23}, 4167-4188 (2006).

\bibitem{Johannsen:2011dh}
T.~Johannsen and D.~Psaltis,
Phys. Rev. D \textbf{83}, 124015 (2011).

\bibitem{Johannsen:2013szh}
T.~Johannsen,
Phys. Rev. D \textbf{88}, no.4, 044002 (2013).

\bibitem{Cardoso:2014rha}
V.~Cardoso, P.~Pani and J.~Rico,
Phys. Rev. D \textbf{89}, 064007 (2014).


\bibitem{Konoplya:2016jvv}
R.~Konoplya, L.~Rezzolla and A.~Zhidenko,
Phys. Rev. D \textbf{93}, no.6, 064015 (2016).

\bibitem{Ghasemi-Nodehi:2016wao}
M.~Ghasemi-Nodehi and C.~Bambi,
Eur. Phys. J. C \textbf{76}, no.5, 290 (2016).

\bibitem{Konoplya:2018arm}
R.~A.~Konoplya, Z.~Stuchl\'\i{}k and A.~Zhidenko,
Phys. Rev. D \textbf{97}, no.8, 084044 (2018).

\bibitem{Chen:2019jbs}
C.~Y.~Chen and P.~Chen,
Phys. Rev. D \textbf{100}, no.10, 104054 (2019).

\bibitem{Carson:2020dez}
Z.~Carson and K.~Yagi,
Phys. Rev. D \textbf{101}, no.8, 084030 (2020).



\bibitem{Cano:2019ore}
P.~A.~Cano and A.~Ruip\'erez,
JHEP \textbf{05}, 189 (2019)
[erratum: JHEP \textbf{03}, 187 (2020)].

\bibitem{Cardoso:2018ptl}
V.~Cardoso, M.~Kimura, A.~Maselli and L.~Senatore,
Phys. Rev. Lett. \textbf{121}, no.25, 251105 (2018).

\bibitem{Cardoso:2019mqo}
V.~Cardoso, M.~Kimura, A.~Maselli, E.~Berti, C.~F.~B.~Macedo and R.~McManus,
Phys. Rev. D \textbf{99}, no.10, 104077 (2019).

\bibitem{Reall:2019sah}
H.~S.~Reall and J.~E.~Santos,
JHEP \textbf{04}, 021 (2019).

\bibitem{McManus:2019ulj}
R.~McManus, E.~Berti, C.~F.~B.~Macedo, M.~Kimura, A.~Maselli and V.~Cardoso,
Phys. Rev. D \textbf{100}, no.4, 044061 (2019).

\bibitem{Sennett:2019bpc}
N.~Sennett, R.~Brito, A.~Buonanno, V.~Gorbenko and L.~Senatore,
Phys. Rev. D \textbf{102}, no.4, 044056 (2020).

\bibitem{Cano:2020cao}
P.~A.~Cano, K.~Fransen and T.~Hertog,
Phys. Rev. D \textbf{102}, no.4, 044047 (2020).

\bibitem{Hatsuda:2020egs}
Y.~Hatsuda and M.~Kimura,
Phys. Rev. D \textbf{102}, no.4, 044032 (2020).


\bibitem{Cunha:2018uzc}
P.~V.~P.~Cunha, C.~A.~R.~Herdeiro and E.~Radu,
Phys. Rev. D \textbf{98}, no.10, 104060 (2018).


\bibitem{Bah:2021jno}
I.~Bah, I.~Bena, P.~Heidmann, Y.~Li and D.~R.~Mayerson,
[arXiv:2104.10686 [hep-th]].

\bibitem{Chen:2020aix}
C.~Y.~Chen,
JCAP \textbf{05}, 040 (2020).



\bibitem{Jefremov:2016dpi}
P.~Jefremov and V.~Perlick,
Class. Quant. Grav. \textbf{33}, no.24, 245014 (2016)
[erratum: Class. Quant. Grav. \textbf{35}, no.17, 179501 (2018)].

\bibitem{Mukherjee:2018dmm}
S.~Mukherjee, S.~Chakraborty and N.~Dadhich,
Eur. Phys. J. C \textbf{79}, no.2, 161 (2019).

\bibitem{Chakraborty:2019rna}
C.~Chakraborty and S.~Bhattacharyya,
JCAP \textbf{05}, 034 (2019).




\bibitem{Destounis:2020kss}
K.~Destounis, A.~G.~Suvorov and K.~D.~Kokkotas,
Phys. Rev. D \textbf{102}, no.6, 064041 (2020).

\bibitem{Destounis:2021mqv}
K.~Destounis, A.~G.~Suvorov and K.~D.~Kokkotas,
Phys. Rev. Lett. \textbf{126}, no.14, 141102 (2021).



\bibitem{Papadopoulos:2018nvd}
G.~O.~Papadopoulos and K.~D.~Kokkotas,
Class. Quant. Grav. \textbf{35}, no.18, 185014 (2018).




\bibitem{Doneva:2021dcc}
D.~D.~Doneva and S.~S.~Yazadjiev,
Phys. Rev. D \textbf{103}, no.8, 083007 (2021).









\bibitem{Rana:2019bsn}
P.~Rana and A.~Mangalam,
Class. Quant. Grav. \textbf{36}, 045009 (2019).


\bibitem{Hughes:1999bq}
S.~A.~Hughes,
Phys. Rev. D \textbf{61}, no.8, 084004 (2000)
[erratum: Phys. Rev. D \textbf{63}, no.4, 049902 (2001); erratum: Phys. Rev. D \textbf{65}, no.6, 069902 (2002); erratum: Phys. Rev. D \textbf{67}, no.8, 089901 (2003); erratum: Phys. Rev. D \textbf{78}, no.10, 109902 (2008); erratum: Phys. Rev. D \textbf{90}, no.10, 109904 (2014)].

\bibitem{Hughes:2001jr}
S.~A.~Hughes,
Phys. Rev. D \textbf{64}, 064004 (2001)
[erratum: Phys. Rev. D \textbf{88}, no.10, 109902 (2013)].

\bibitem{Fayos:2007ks}
F.~Fayos and C.~Teijon,
Gen. Rel. Grav. \textbf{40}, 2433-2460 (2008).

\bibitem{Teo:2020sey}
E.~Teo,
Gen. Rel. Grav. \textbf{53}, no.1, 10 (2021).


\bibitem{EventHorizonTelescope:2019dse}
K.~Akiyama \textit{et al.} [Event Horizon Telescope],
Astrophys. J. Lett. \textbf{875}, L1 (2019).

\bibitem{EventHorizonTelescope:2020qrl}
D.~Psaltis \textit{et al.} [Event Horizon Telescope],
Phys. Rev. Lett. \textbf{125}, no.14, 141104 (2020).

\bibitem{Brahma:2020eos}
S.~Brahma, C.~Y.~Chen and D.~h.~Yeom,
Phys. Rev. Lett. \textbf{126}, no.18, 181301 (2021).

\bibitem{Shaikh:2021yux}
R.~Shaikh, K.~Pal, K.~Pal and T.~Sarkar,
Mon. Not. Roy. Astron. Soc. \textbf{506}, no.1, 1229-1236 (2021).

\bibitem{EventHorizonTelescope:2021dqv}
P.~Kocherlakota \textit{et al.} [Event Horizon Telescope],
Phys. Rev. D \textbf{103}, no.10, 104047 (2021).

\bibitem{Zakharov:2021gbg}
A.~F.~Zakharov,
[arXiv:2108.01533 [gr-qc]].






\bibitem{Gralla:2020pra}
S.~E.~Gralla,
Phys. Rev. D \textbf{103}, no.2, 024023 (2021).

\bibitem{Volkel:2020xlc}
S.~H.~V\"olkel, E.~Barausse, N.~Franchini and A.~E.~Broderick,
[arXiv:2011.06812 [gr-qc]].

\bibitem{Glampedakis:2021oie}
K.~Glampedakis and G.~Pappas,
[arXiv:2102.13573 [gr-qc]].

\bibitem{Nampalliwar:2021oqr}
S.~Nampalliwar and S.~K,
[arXiv:2108.01190 [gr-qc]].


\bibitem{Lima:2021las}
H.~C.~D.~Lima, Junior., L.~C.~B.~Crispino, P.~V.~P.~Cunha and C.~A.~R.~Herdeiro,
Phys. Rev. D \textbf{103}, no.8, 084040 (2021).


\bibitem{Page:1974he}
D.~N.~Page and K.~S.~Thorne,
Astrophys. J. \textbf{191}, 499-506 (1974).

\bibitem{Thorne:1974ve}
K.~S.~Thorne,
Astrophys. J. \textbf{191}, 507-520 (1974).



\bibitem{Reynolds:2019uxi}
C.~S.~Reynolds,
Nature Astron. \textbf{3}, no.1, 41-47 (2019).


\bibitem{Dabrowski:2000qv}
Y.~Dabrowski and A.~N.~Lasenby,
Mon. Not. Roy. Astron. Soc. \textbf{321}, 605 (2001).

\bibitem{Mueller:2003fr}
A.~Mueller and M.~Camenzind,
Astron. Astrophys. \textbf{413}, 861-878 (2004).

\bibitem{Mueller:2006dn}
A.~Mueller and M.~Wold,
Astron. Astrophys. \textbf{457}, 485-492 (2006).

\bibitem{Bambi:2017khi}
C.~Bambi,
\textit{Black Holes: A Laboratory for Testing Strong Gravity}, Springer (2017).

\bibitem{Bini:2001kx}
D.~Bini, R.~T.~Jantzen and B.~Mashhoon,
Class. Quant. Grav. \textbf{19}, 17-38 (2002).

\bibitem{Datta:2020axm}
S.~Datta and S.~Mukherjee,
Phys. Rev. D \textbf{103}, no.10, 104032 (2021).











\end{thebibliography}
\end{document}